\documentclass[aps,pre,preprint,groupedaddress,showpacs]{revtex4-1}
\usepackage{graphicx,amsmath}
\usepackage{caption}
\begin{document}

\title{Heat Conduction in a hard disc system with non-conserved momentum}


\author{P.L. Garrido}
\email[]{garrido@onsager.ugr.es}
\affiliation{Instituto Carlos I de F{\'\i}sica Te{\'o}rica y Computacional. Universidad de Granada. E-18071 Granada. Spain }

\author{Joel L. Lebowitz}
\email[]{lebowitz@math.rutgers.edu}
\affiliation{Department of Mathematics and Physics, Rutgers University, Piscataway, New Jersey 08854, USA }

\date{\today}
\begin{abstract}
We describe results of computer simulations of steady state heat transport in a fluid of hard discs undergoing both elastic interparticle collisions and velocity randomizing collisions  which do not conserve momentum. The system consists of $N$ discs of radius $r$ in a unit square, periodic in the y-direction and having thermal walls at $x=0$ with temperature $T_0$ taking values from $1$ to $20$  and  at $x=1$ with $T_1=1$. We consider different values of the ratio between randomizing and interparticle collision rates and extrapolate results from different $N$, to $N\rightarrow\infty$, $r\rightarrow 0$ such that $\rho=\pi r^2 N=1/2$. We find that in the (extrapolated) limit $N\rightarrow\infty$, the systems local density and temperature profiles are those of local thermodynamic equilibrium (LTE) and obey Fourier's law. The variance of global quantities, such as the total energy, deviates from its local equilibrium value in a form consistent with macroscopic fluctuation theory.
\end{abstract}
\pacs{18-3e}
\maketitle

\section{Introduction}

We continue our investigation, via molecular dynamics (MD), of the nonequilibrium stationary states of a system of hard discs of radius $r$ in a unit square \cite{GL}. The system  has periodic boundary conditions in the $y$-direction and  thermal walls at $x=0$ and $x=1$ with temperatures $T_0$ and $T_1$($=1$) respectively. The areal density $\rho=N\pi r^2=0.5$  for all computer simulated cases in this paper. We have simulated different $N$ values ranging from $460$ up to $5935$ (see Appendix I for all the technical details about the computer simulation) in order to do a finite size analysis  and obtain the hydrodynamic description of the system, in the limit, $N\rightarrow\infty$, $r\rightarrow 0$.

The dynamics of the discs consists in linear displacements at constant velocity and elastic collisions when two discs meet. Additionally  the dynamics has a part that breaks the bulk momentum conservation of the system dynamics. 
 We introduced such a mechanism recently in the context of kinetic equations for the one particle distribution, $f(r,v,t)$\cite{GL}. There we added to the usual collision term, $Q_c(f)$, such as Boltzmann, Boltzmann-Enskog and BGK, a linear collision term, $Q_D(f)$,  which randomizes velocities but conserves energy. We multiplied this term by a parameter $\alpha$,
 \begin{equation}
\partial_t f+v\cdot\nabla f=Q_C(f)+\alpha Q_D(f)
\end{equation} 
 This led to an evolution of $f$ which had only two conservation laws, particle and energy density, i.e. no momentum conservation. $Q_D(f)$ represents particle collisions with fixed obstacles, as in a Lorentz gas, corresponds physically to a fluid moving in a porous medium \cite{Callen}.
Diffusively scaling space and time \cite{S} enabled us to derive rigorously, from the modified kinetic equations, macroscopic equations for the two conserved quantities  \cite{EGLM}. We are not aware of such rigorous derivation for the full conservation laws including momentum.

In this note we describe MD simulations of hard discs with a dynamics which destroys momentum conservation in a   different way from that modeled in Ref.\cite{GL} but has physically a similar effect. We do not assume here the validity of any kinetic equation for $f$. Following each elastic collision between a pair of particles in the system we randomize the direction of $\gamma$ particles ($\gamma\in[0,10]$) chosen at random (excluding the particles involved in the actual collision to exclude dynamic pathologies). We have simulated $\gamma=0.3, 1,5$ and $10$. 

In addition to these bulk collisions there are also collisions with the thermal walls.
When a disc hits a thermal wall it gets a new normal component of the velocity with respect to the boundary. Its value is obtained from a Maxwellian velocity distribution with the temperature that corresponds to the wall. 
We use this combined dynamics to the study of the stationary state. This was previously investigated for the case $\gamma=0$ \cite{delPozo}. We have done simulations  for $T_0=1,3,5,7,\ldots,21$.

 We find that for $N\rightarrow\infty$  our system satisfies locally the equilibrium equation of state, which  is independent of the $\gamma$ parameter and the heat transport satisfies Fourier's Law so that 
\begin{subequations}\label{leq}
\begin{align}
Q&=T(x)\rho(x)\bar H(\rho(x))\label{leq1}\\
J&=-\kappa(T,\rho)\frac{dT}{dx} \quad\quad x\in[0,1] \label{leq2}
\end{align}
\end{subequations} 
Here $Q$ is the reduced pressure ($Q=P\pi r^2$, with $P$ the pressure), $J$ is the reduced heat current ($J=J_H/r$, with $J_H$ the heat current) and $\rho(x), T(x)$ are the {\it local areal density} and the {\it local temperature} respectively. The boundary conditions, in the limit $N\rightarrow\infty$,are
\begin{equation}
T(0)=T_0\quad,\quad T(1)=1\quad,\quad \bar\rho=\int_{0}^1dx\,\rho(x)=0.5
\end{equation}
Both $P$ and $J$ are constant in the stationary state. 

To obtain expressions for $\rho(x)$ and $T(x)$  we need to know $\bar H$ and $\kappa$. For $\bar H$ we use  Henderson's hard discs equation of state \cite{Henderson}  known to be a very good approximation for $\rho\lesssim 0.6$:
\begin{equation}
\bar H(\rho)=\frac{1+\rho^2/8}{(1-\rho)^2}-0.043\frac{\rho^4}{(1-\rho)^3}\label{Hen}
\end{equation}

To obtain an expression for $\kappa(T,\rho)$ we first note that to the extent that (\ref{leq}) holds, i.e. $J$ is a linear functional of the local gradient, $J$ will be proportional to $\sqrt{T}$. It will vanish as $r\rightarrow 0$ so it has to be rescaled by $r$ as is done in (\ref{leq2}). Thus, in the limit $N\rightarrow\infty$
\begin{equation}
\kappa(T,\rho)=\sqrt{T}K(\rho;\gamma)
\end{equation}

To find $K(\rho;\gamma)$ we use an expression for the conductivity derived in Ref.\cite{GL} from an approximate solution of the Enskog equation with a momentum destroying collision term proportional to $\alpha$, a.f. eq. (78) in \cite{GL}. We replace the dimensionless parameter $\alpha_s$ there by a function $\phi(\gamma)$. This gives
\begin{equation}
K(\rho;\gamma)=\frac{1+(3+16 \phi(\gamma)\rho)\rho\chi(\rho)+(19+9\pi)(\rho\chi(\rho))^2/4\pi}{\sqrt{\pi}(\chi(\rho)+4\pi \phi(\gamma))}\label{th}
\end{equation}
where $\chi(\rho)=(H(\rho)-1)/2\rho$. $\phi(\gamma)$ is chosen to give accurate results at low densities. This gives, for the values of $\gamma$ simulated here, $\phi(0.3)=0.09$, $\phi(1)=0.286$, $\phi(5)=1.685$ and $\phi(10)=5.50$. Using these functions for $Q$ and $K$ we can get the profiles $T(x)$ 
 and $\rho(x)$ for any given $T_0$. This is what we do in the remainder of the paper.

\section{Local equilibrium}

The local equilibrium hypothesis used in the last section assumes that in  the hydrodynamic description of a macroscopic system we can locally define equilibrium thermodynamic observables that obey the equilibrium relations between them. To check if the equilibrium equation of state (EOS) holds in each stripe parallel to the y-axis we define the local density for the stripe $s$ at time $t$, $\rho(s,t)$ as  the number of particle centers in the stripe:
\begin{equation}
\rho(s;t)=\frac{1}{N}\sum_{i:r_i(t)\in B(s)}1
\end{equation}
(see Appendix).
We then use the virial theorem to compute the local pressure $Q_v(s)$ :
 \begin{equation}
 Q_v(s)=\rho(s) T(s)+\frac{1}{2 \Delta_B^2 \tau}\sum_{n:t_n\in[0,\tau];r_i(t)\in B(s)}r_{ij}(t_n)\cdot p_{ij}(t_n)
 \label{prev}
 \end{equation}
where $r_{ij}=r_i-r_j$, $p_{ij}=p_i-p_j$, $\Delta_B$ is the width of a stripe, and the sum runs over all particle-particle collisions that occur in the cell $s$ in the time interval $[0,\tau]$ letting $\tau$ be large enough for the right hand side of eq. (\ref{prev}) to be independent of $\tau$.   

First we checked that the {\it virial pressure} $Q_v$ is constant all over the system for each $N$, temperature gradient and $\gamma$'s and we got the average over the stripes. Finally from such data  we extrapolated its value to $N\rightarrow\infty$. We also checked that this value agrees with the pressure measured at the thermal walls, by computing the momentum transfer to the wall.

\begin{figure}[h!]
\begin{center}
\includegraphics[width=7cm]{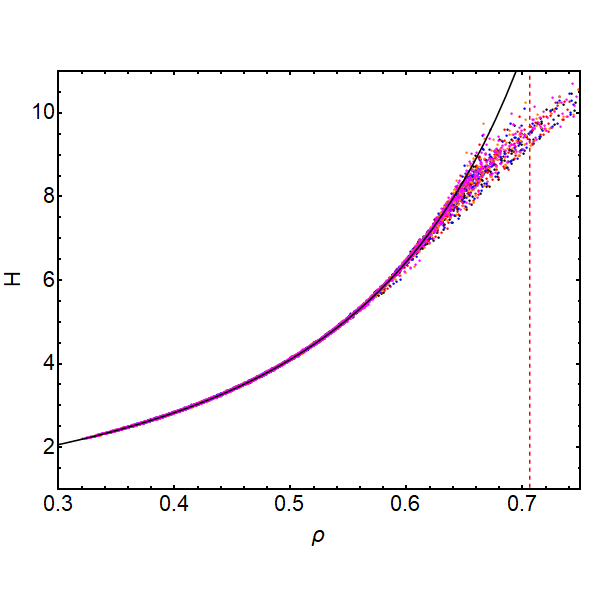}
\includegraphics[width=7cm]{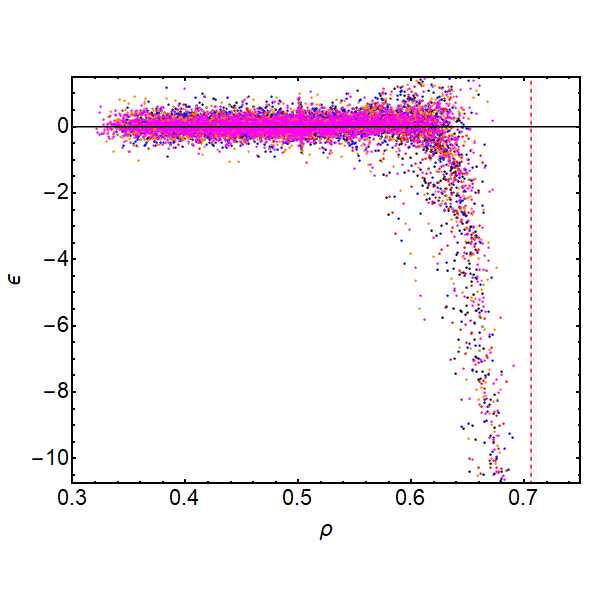}   
\vskip -0.5cm
\caption{\small\it\raggedright Left: $H(s)=Q_v(s)/\rho(s)T(s)$ vs. $\rho(s)$ for $s=4,\ldots,N_B-3$  for all the simulated $N$'s in table \ref{TableI} and $\Delta T=T_0-T_1=0$, $2$, $\ldots$, $20$. Black, Red, Blue, Orange  and Magenta points are  for $\gamma=0$, $0.3$, $1$, $5$ and $10$ respectively. The dotted vertical line is the observed equilibrium phase transition critical value ($\gamma=0$ and $\Delta T=0$): $\rho_c\simeq 0.7062$. The black solid line is the $\bar H(\rho)$  function corresponding to the Henderson's equation of state (see text). Right: Relative error between simulated results for $H(s)$ and the Henderson's $\bar H(\rho)$ proposal. \label{eoss}}
\end{center}
\end{figure}

In order to check the local equilibrium hypothesis we plot $H(s)= Q_v(s)/\rho(s)T(s)$ vs $\rho(s)$ for all the stripes $s$ (except the ones near the heat bath boundaries) and for all the simulations done: $\gamma$'s, $\Delta T$'s and $N$'s. In total there are $17195$ data points represented in figure \ref{eoss}.
We observe in figure \ref{eoss} that most of the points with local densities $\lesssim 0.6$ follow the same curve independently of the position of the stripe, the number of particles $N$, the temperature gradient $\Delta T$ or the randomization intensity $\gamma$. 
We observe that Henderson's EOS (\ref{Hen}) follows the data almost perfectly for low densities and only deviates when approaching the expected phase transition critical density at $\rho_c\simeq 0.7062$. Moreover, the data lose the scaling property for densities larger than $0.6$. That is due mainly to the small size of the virtual stripes when the particles begin to crystallize in a hexagonal lattice and the center of the particles tend to be ordered and aligned. Therefore nearby  stripes may contain one or two lines of centers affecting microscopically  the  values of the measured densities. Finally we can conclude from the data analysis that our nonequilibrium system has the local equilibrium property.

The quality of the data allows us to look with detail the quality of the Henderson EOS. We have plotted in figure \ref{eoss} the relative error between the data and the proposal: 
\begin{equation}
\epsilon(s)=100 \frac{H(s)-\bar H(\rho(s))}{H(s)}
\end{equation}
We observe that for $\rho\leq 0.6$ the  Henderson's EOS has less than $1\%$ of relative error with the data.  

\section{Temperature Profiles and Fourier's Law}

The local temperature is defined as the average local kinetic energy per particle at each stripe. That is, 
\begin{equation}
T(s)=\frac{1}{2N(s)M}\sum_{t=1}^M\sum_{i:r_i(t)\in B(s)}v_i(t)^2 \quad,\quad N(s)=\frac{1}{M}\sum_{t=1}^M\sum_{i:r_i(t)\in B(s)}1\label{temp}
\end{equation}
where $B(s)$ is the set of point belonging to the stripe $s$. We assume that a particle belongs to a stripe if its center is in the stripe. This computational method is efficient but it does not compute correctly the density behavior near the walls.  

\begin{figure}[h!]
\begin{center}
\includegraphics[width=8cm]{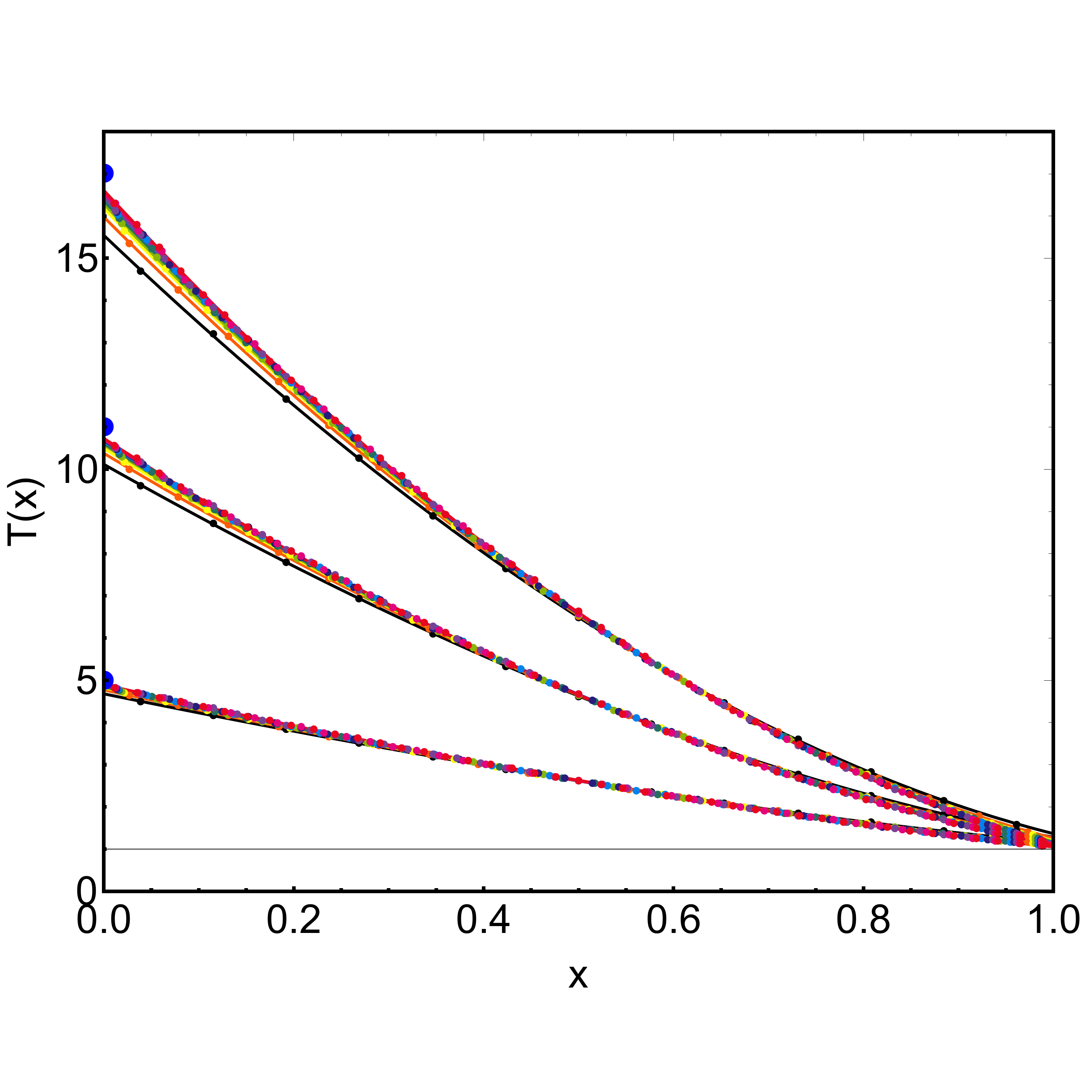}   
\includegraphics[width=8cm]{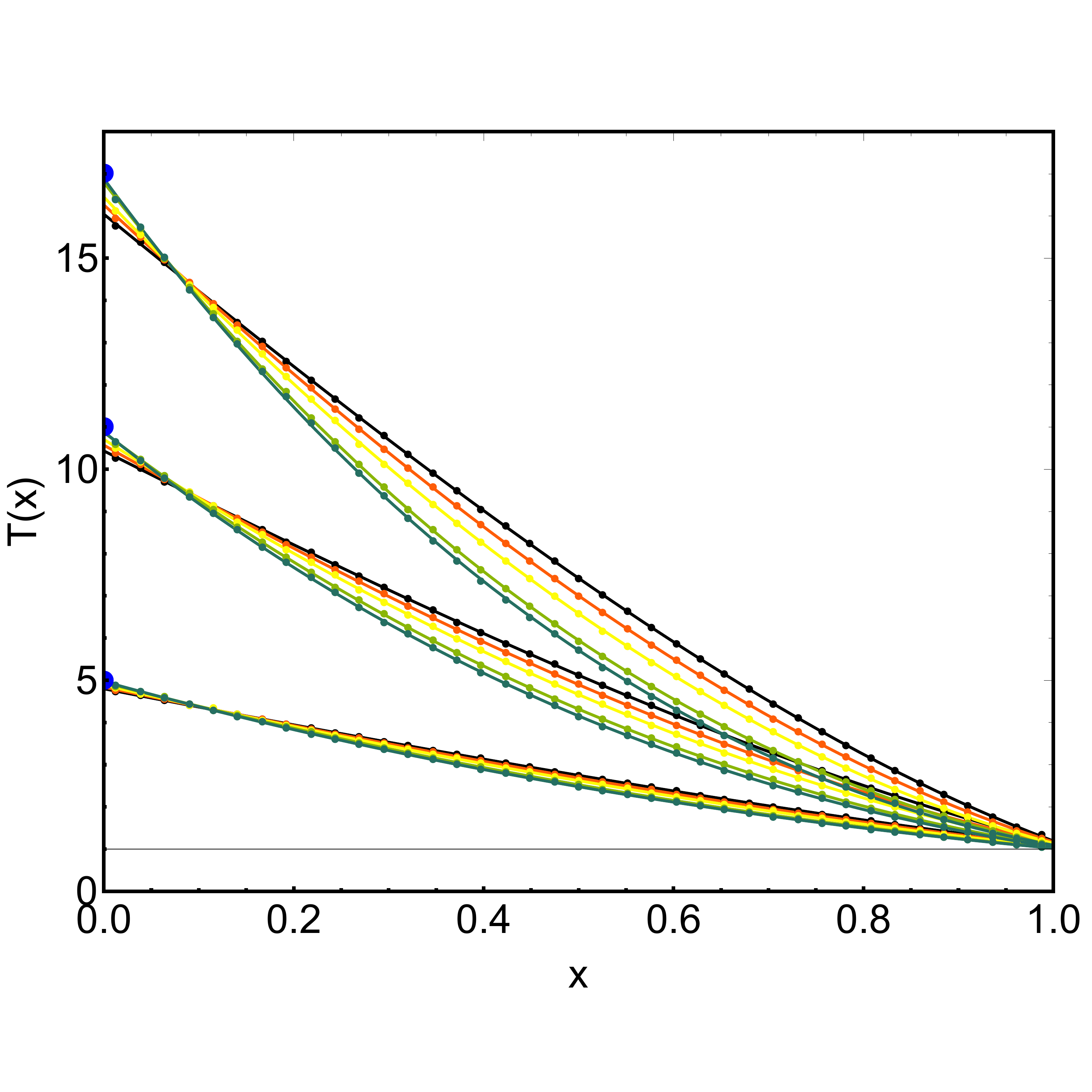}
\vskip -0.5cm
\caption{\small\it\raggedright Left: Thermal profiles for all the simulated $N$'s in table \ref{TableI} and $\Delta T=T_0-T_1=4$, $10$, $16$ and $\gamma=1$. Each visual group of data corresponds to the same $\Delta T$ and different colors to different $N$ values (for instance black dots correspond to the smallest $N$ and orange dots to the largest one).  Right: Thermal profiles for $N=4000$, $\Delta T=4$, $10$, $16$ and $\gamma=0$ (Dark Green), $\gamma=0.3$ (Green), $\gamma=1$ (Yellow), $\gamma=5$ (Red) and $\gamma=10$ (Black).
Solid lines are the corresponding fitted functions. Big points at $x=0$ show the simulated $T_0$ values. Error bars for each data point are shown.
\label{fou1}}
\end{center}
\end{figure}

As an example, we  show in figure \ref{fou1} the temperature profiles  for  $\Delta T=4$, $10$ and $16$. In  figure   \ref{fou1}a  we see the case $\gamma=1$ and we show the size effect in the profiles. In figure \ref{fou1}b  we show the effect of $\gamma$ for  $N=4000$.  We see that all the measured profiles are monotone decreasing functions with positive curvature. The size effects are larger as we increase the temperature gradient for a given $\gamma$ value. These all follows from the form of $\bar H$ and $K$ given in (\ref{Hen}) and (\ref{th}).

We observe that the extrapolated profiles up to the boundaries do not coincide with the temperature values used in the simulations.  This phenomena is known as {\it thermal resistance}. Kinetic theory arguments predict that  this temperature gap goes to  zero as the mean free path which behaves like $N^{-1/2}$ when $N\rightarrow\infty$. We have checked this prediction by fitting the data for each temperature profile  to a polynomial with positive curvature. We extrapolate the fitted functions to the boundary points $x=0$ and $x=1$ getting $T_0^e$ and $T_1^e$ respectively. We define 
the relative gap:
\begin{equation}
G_i(N)=\frac{(T_i-T_i^e(N))}{T_i}\simeq \frac{G_i}{\sqrt{N}}\quad (i=0,1)
\end{equation} 
We got the values of $G_i$ versus $\Delta T$ by extrapolating  $G_{0,1}(N)\sqrt{N}$  to $N\rightarrow\infty$ for the different values of $\gamma$. We confirm the behavior with $N^{-1/2}$ of $G_i(N)$. We see that for a given temperature gradient, and size $N$ the gap increases with $\gamma$. On the other hand, for any $\gamma$ value,  the thermal resistance increases with the gradient.

\begin{figure}[h!]
\begin{center}
\includegraphics[width=9cm]{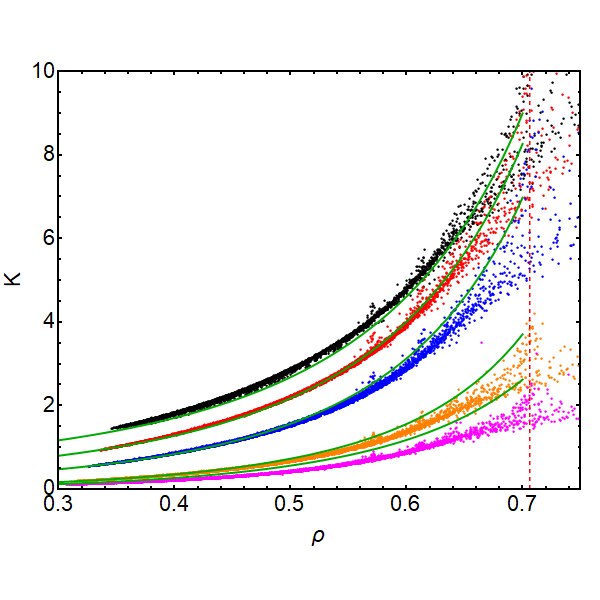}   
\vskip -0.5cm
\caption{\small\it\raggedright $K(s)$ vs. $\rho(s)$ for the virtual boxes $s=2,\ldots,N_B-1$  for all the simulated $N$'s in table \ref{TableI} and $\Delta T=T_0-T_1=0$, $2$, $\ldots$, $20$. Black, Red, Blue, Orange and Magenta dots are for  $\gamma=0$, $0.3$, $1$, $5$ and $10$ respectively. The Red vertical line is the observed equilibrium phase transition critical value when $\gamma=0$ and $\Delta T=0$: $\rho_c\simeq 0.7062$. The solid lines are obtained by fitting the parameters $\phi(\gamma)$ in the Boltzmann-Enskog expression for the thermal conductivity  (see eq. \ref{th}) to the data and by using the Henderson's EOS.
\label{fouu2}}
\end{center}
\end{figure}

To check the validity of Fourier's law (\ref{leq}b) we computed at each stripe
\begin{equation}
K(\rho;\gamma)=-\frac{J}{\sqrt{T(x)}}\left(\frac{dT}{dx}\biggr\vert_{x=x(\rho)}\right)^{-1}
\end{equation}
as a function of $\rho$.
If Fourier's law holds $K(\rho;\gamma)$ should be, for each $\gamma$, an universal curve independent of the parameters that define the stationary state: $T_0$, $T_1$ and $\bar\rho$ and the $x$ used. The derivative of the thermal profile is analytically done over the fitted profile.  The use of fitting functions with positive curvature reduces the dispersion of the values of the derivatives  due to the typical ``waves'' around the average profile that one obtains when using an arbitrary polynomial.   For the thermal conductivity analysis we have discarded the points near the thermal walls  to minimize the boundary effects.

We draw figure \ref{fouu2} by computing the derivative of the fitted function at the center of each virtual box, associating to the point the measured average density in the box. We see there that the data follows quite well a unique curve for each given $\gamma$ value. Let us stress that for each $\gamma$ we are plotting around $3600$ data points corresponding to systems with different thermal gradients, number of particles in  each of  the stripes where local variables are defined. Observe that the $N$ dependence is not visible and that the $K$ decreases with $\gamma$ as expected. We observe deviations to the unique curve when approaching the critical density.   As we see, we have obtained a reasonable description by $K(\rho)$ of eq. (\ref{th}) for $\rho\leq 0.6$. Again the deviations from the theory to the data increase with $\gamma$.

\section{Fluctuations}

\begin{figure}[h!]
\begin{center}
\includegraphics[width=5cm]{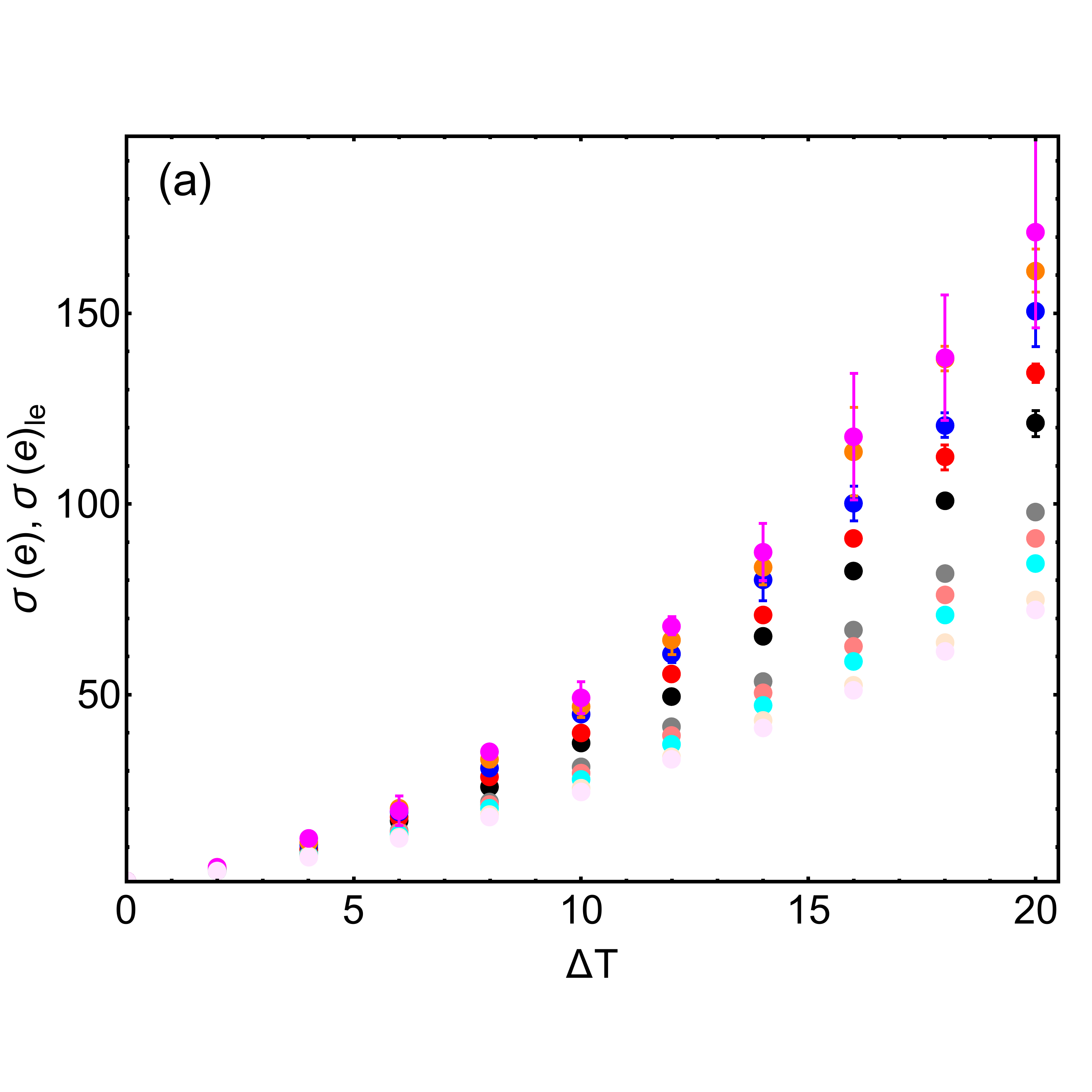}
\includegraphics[width=5cm]{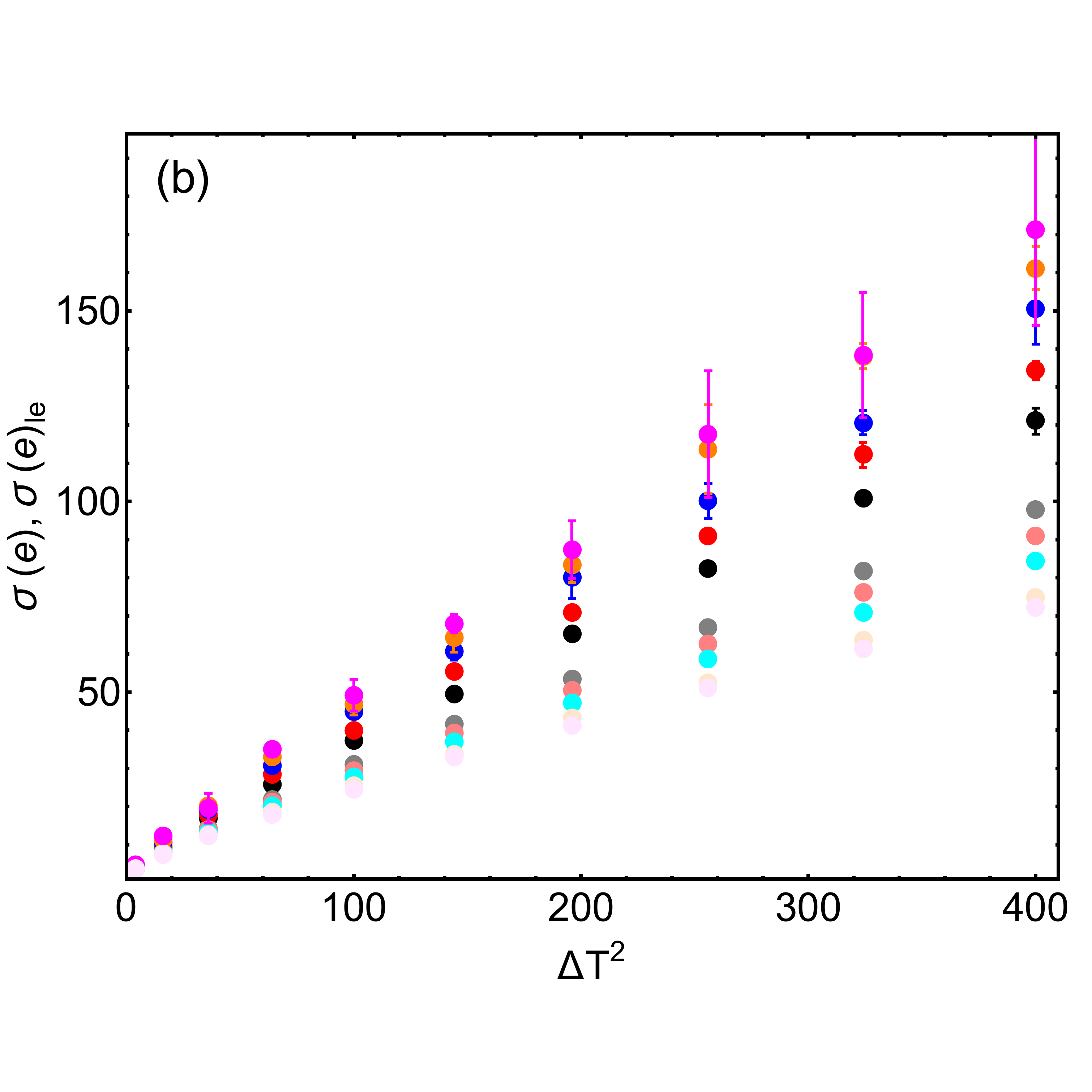}
\includegraphics[width=5cm]{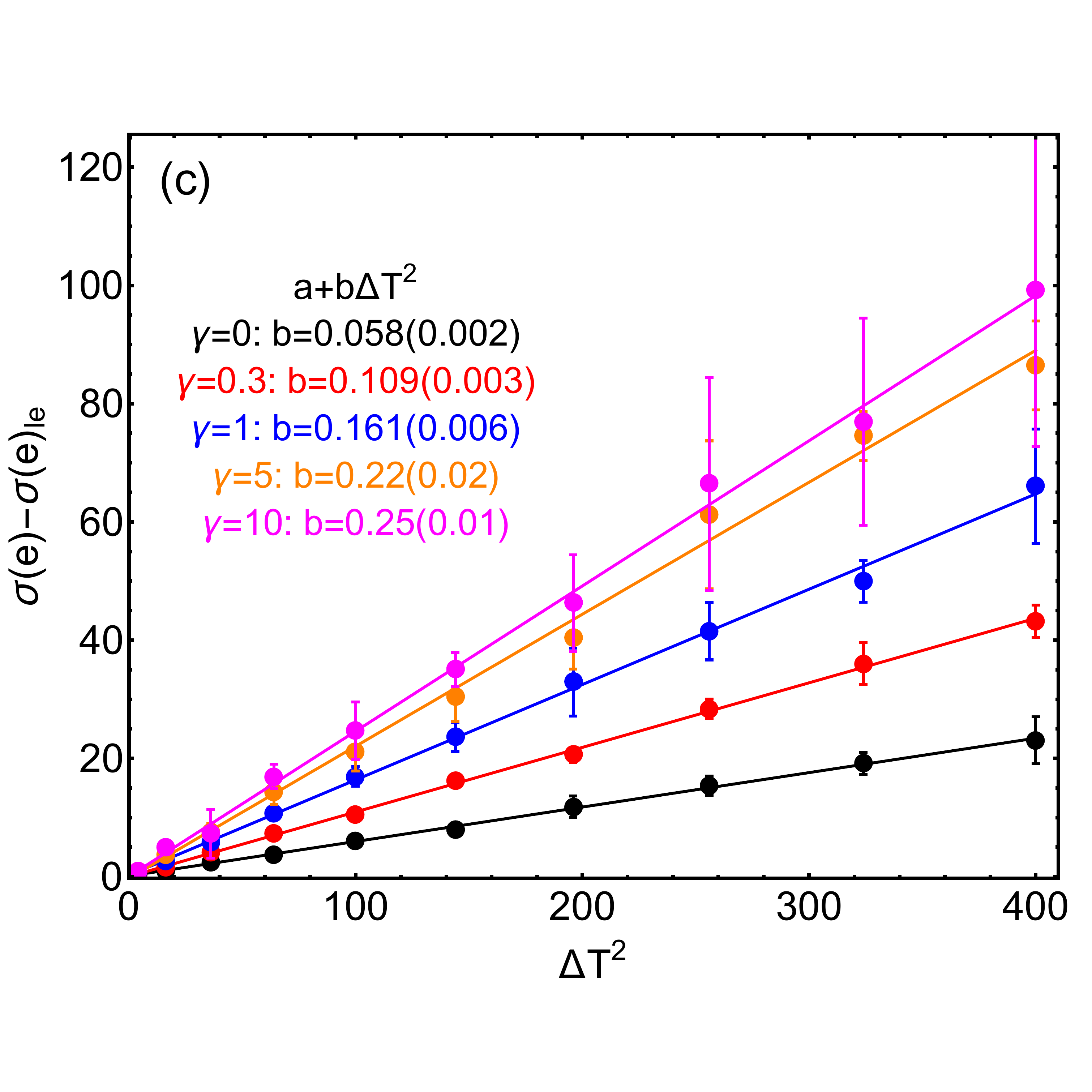}   
\vskip -0.5cm
\caption{\small\it\raggedright (a): Energy fluctuations $\sigma(e)$ vs $\Delta T$ for $\gamma=0$ (black points), $0.3$ (red points),  $1$ (blue points), $5$ (orange points) and $10$ (magenta points). We also plot the energy fluctuations assuming local equilibrium, $\sigma(e)_{le}$, for $\gamma=0$ (gray points), $\gamma=0.3$ (pink points), $\gamma=1$ (cyan points), $\gamma=5$ (light orange points) and $\gamma=10$ (light magenta points). (b): Same data as left figure but plotted versus $\Delta T^2$ to see the nontrivial curvature of the data sets. (c): Difference between the measured energy fluctuations and the local equilibrium energy fluctuations. Solid lines are linear fits to the data.  \label{m2N11}}
\end{center}
\end{figure}

We  measured the fluctuations of the energy per particle in the stationary state:
\begin{equation}
m(e_N)=\langle e_N^2\rangle-\langle e_N\rangle^2\quad ,\quad e_N=\frac{1}{N}\sum_{i=1}^N e(v_i)
\end{equation}
this has the asymptotic behavior:
\begin{equation}
\sigma(e)=\lim_{N\rightarrow\infty}Nm(e_N)
\end{equation}
plotted in Fig. \ref{m2N11}a.

It is well known, for certain exactly solvable models, e.g. SEP, KMP that even when the system is in LTE in the macroscopic limit ($N\rightarrow\infty$) the fluctuation in global quantities deviates from their LTE values \cite{DLS}.
For our system
\begin{equation}
\sigma_{le}(e)=\rho^{-1}\int_0^1dx\,\rho(x)T(x)^2
\end{equation}
This is plotted in Figure  \ref{m2N11}b.

In Figure \ref{m2N11}c we show $\delta(e)\equiv\sigma(e)-\sigma_{leq}(e)$ and observe that $\delta(e)=b(\gamma)(\Delta T)^2$ with $b(\gamma)>0$ monotone increasing with $\gamma$:  $b(0)=0.058(0.002)$, $b(0.3)=0.109(0.003)$, $b(1)=0.161(0.006)$, $b(5)=0.22(0.02)$ and $b(10)=0.25(0.01)$. This is of the same form as that founded for the exactly solvable models. It is also of the form found by macroscopic fluctuation theory (MFT) \cite{Bertini}. The coefficient of $(\Delta T)^2$ can be related in MFT (when there is only a single macroscopic variable) to the compressibility and transport coefficients. It can be positive or negative. We have not tried to compute $b(\gamma)$ for our system but it is noteworthy that the linear $(\Delta T)^2$ dependence holds also for deterministic systems with $\gamma=0$.

\section{Concluding Remarks}

We have studied via MD simulations the NESS in the scaling limit, $r\rightarrow 0$ of a system of hard discs with a mechanism that breaks the conservation of momentum. The results are consistent with the system being in LTE. We also found evidence of long range correlations behaving as $N^{-1}$ giving rise to non LTE variances in global quantities.

\section{Acknowledgements}
We thank H. Spohn, C. Bernardin and specially R. Esposito and D. Gabrielli for very helpful correspondences. This work was supported in part by AFOSR [grant FA-9550-16-1-0037].
 PLG was supported also by the Spanish government project FIS2017-84256-P funded by MINECO/AEI/FEDER. We thank the IAS System Biology divison for its hospitality during part of this work.

\section*{Appendix}

Starting in an  ordered initial configuration we let the system  relax towards its stationary state during $5\times 10^4N$ particle collisions and then we do measurements each $10^2N$ particle collisions. The errors in all magnitudes are $3\sigma$ with $\sigma$ being the standard deviation of the set of measurements.
We have measured global magnitudes as the energy per particle, $e_N$,  the pressure, $Q_N$ and the heat current $J_N$ that are defined for the simulations:
\begin{eqnarray}
e_N&=&\frac{1}{N_{meas}}\sum_{n=1}^{N_{meas}}\frac{1}{N}\sum_{i=1}^N e(v_i) \nonumber\\
Q_N&=&\frac{\pi r^2}{\tau_{col}}\sum_{n=1}^{N_{wc}}\left(\Delta v_x\right)_n \nonumber\\
J_N&=&\frac{r}{\tau_{col}}\sum_{n=1}^{N_{wc}}\left(\Delta e(v)\right)_{n}\label{mac}
\end{eqnarray}
where
\begin{equation}
e(v)=\frac{1}{2}(v_{i,x}^2+v_{i,y}^2)
\end{equation}
is the kinetic energy of particle $i$. $N_{meas}$ is the number of measurements done, $N_{wc}$ is the number of hard discs collisions with the walls in the time interval $[0,\tau_{col}]$ once the systems reaches the stationary state. $\left(\Delta A\right)_n$ is the variation of the magnitude $A$ before and after the $n$thcollision with the wall.  

 In order to measure local observables such as the density, temperature and the virial pressure.
 we divide the system  into virtual  stripes parallel to the heat bath walls.  We choose their width to be of order $4r$. More precisely, the number of stripes is $N_B=int(1/2r)$ and the stripes width $l=1/N_B$. Observe that $N_B$ depends on the particle radius that also depends on $N$ as  is seen in Table \ref{TableI}.
\begin{table}[h]
  \centering
\begin{tabular}{| c c|| c c|}
  \hline
 $N$&$N_B$ & $N$ &$N_B$\\
\hline
  460 & 13 & 3434 & 36 \\
  941 & 19 & 3886 & 39 \\
  1456 & 23 &  4367 & 41 \\
 1927 & 27 & 4875 & 43 \\
  2438 & 30 &  5412 & 46 \\
  2900 & 33 &  5935 & 48 \\
    \hline
\end{tabular}
  \caption{$N$ is the different number of particles in the bulk simulated in this paper. $N_B$ is the number of virtual stripes in which we divide the system to measure local observables.}\label{TableI}
\end{table}

\begin{figure}[h!]
\begin{center}
\begin{center}
\includegraphics[width=8cm]{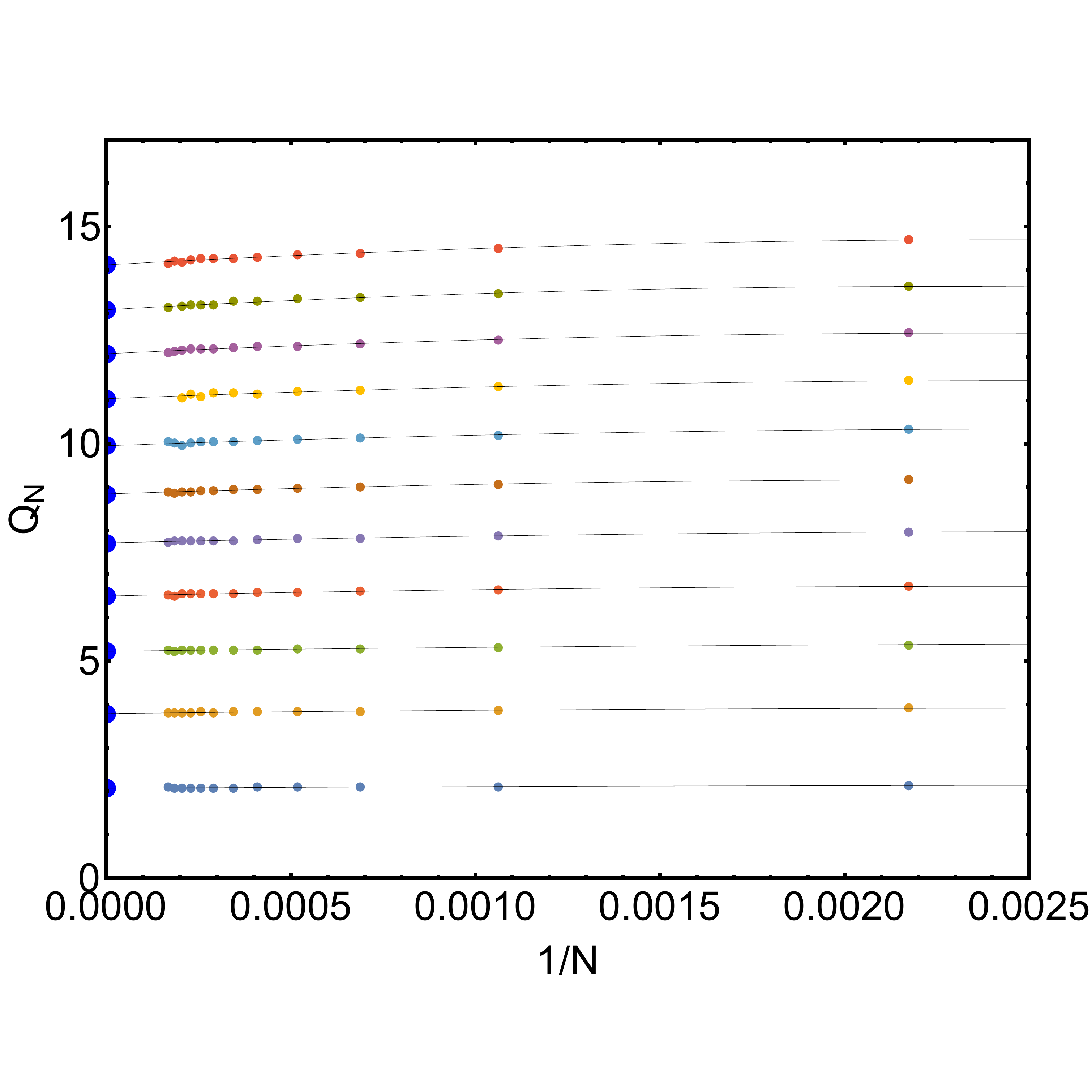}   
\end{center}
\vskip -1cm
\caption{Pressure, $Q_N$, for $\gamma=1$ and  $\Delta T=T_0-T_1=0$, $2$, $\ldots$, $20$ (from bottom to top inside each graph) as a function of $1/N$.  The points on the $1/N=0$ axis correpond to the extrapolation of a parabolic least square fit to the data (solid lines) for each $T_0$. Error bars are smaller than the points size.\label{eN_al}}
\end{center}
\end{figure}

\begin{figure}[h!]
\begin{center}
\includegraphics[width=7cm]{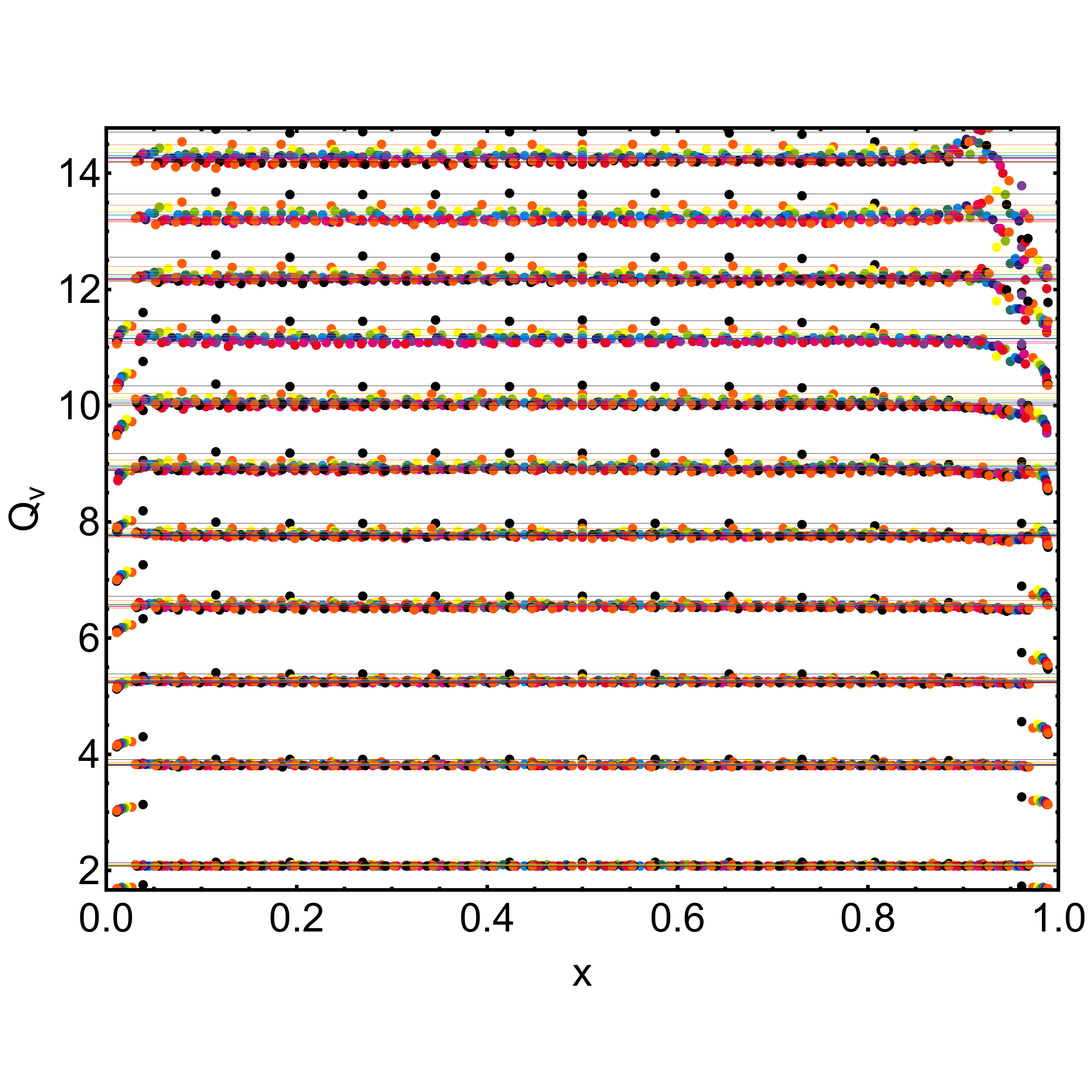}   
\vskip -0.5cm
\caption{\small\it\raggedright Virial pressure for $\gamma=1$ measured at each stripe for $\Delta T=T_0-T_1=0$, $2$, $\ldots$, $20$ (from bottom to top inside each graph). Each color indicate a size.  Thin lines are linear fits of the data (excluding two points near to each boundary) for a given $N$ and $\Delta T$.\label{pree1}}
\end{center}
\end{figure}

For each averaged observable we find a systematic dependence on $N$ and in order to check the hydrodynamic equations we need to extrapolate to their limit, $N\rightarrow\infty$ value. For instance we show in figure \ref{eN_al} the pressure measured at the walls, $Q_N$, as an example of such  systematic size dependence. We plot there  $Q_N$  for $\gamma=1$. In order to extrapolate the data for a given $\Delta T$ and $\gamma$ to the $N\rightarrow\infty$ limit,  it is typically enough to use a second order polynomial fitting function: $a_N=a+a_1/N+a_2/N^2$.
In figure \ref{eN_al} we plot the fits as thin dashed lines. The big points in the figures are the extrapolated $Q$'s  from the fitted functions for each $\Delta T$ values. These limiting values are essential to get a coherent picture of the system's behavior. For instance, in the pressure case, we also measured the local pressure by using the virial expression.  In figure \ref{pree1} we show the measured values of $Q_v(s)$, $s=1,\ldots,N_B$ for a given $\gamma$, $\Delta T$ and $N$.
First we observe that the data at the boundaries are distorted by the way we have defined the density measurements. We discard such boundary data in the analysis. Second, we tried to fit several functions to the bulk data and we concluded that the best one is the constant function. The virial pressure is derived from classical mechanics  and therefore its constant value along the stripes indicates a kind of mechanical equilibrium at the system stationary state. We again use the fitted values for different $N$'s to  extrapolate the data to $N\rightarrow\infty$ by using a second order polynomial in $1/N$. Finally, we can compare these asymptotic values with the pressure $Q$ measured on the walls  and we get a very good match. 

\begin{figure}[h!]
\begin{center}
\includegraphics[width=8cm]{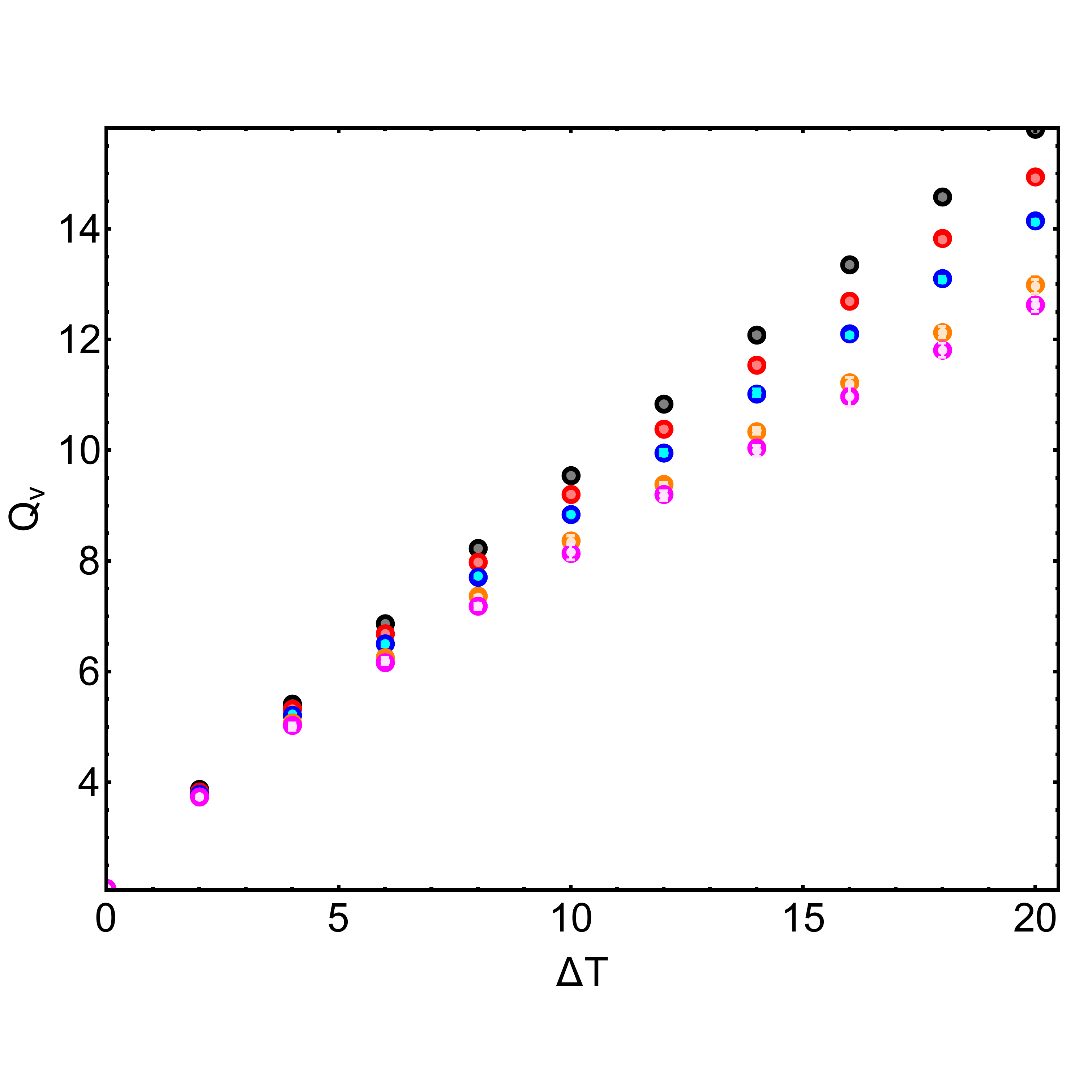}
\includegraphics[width=8.1cm]{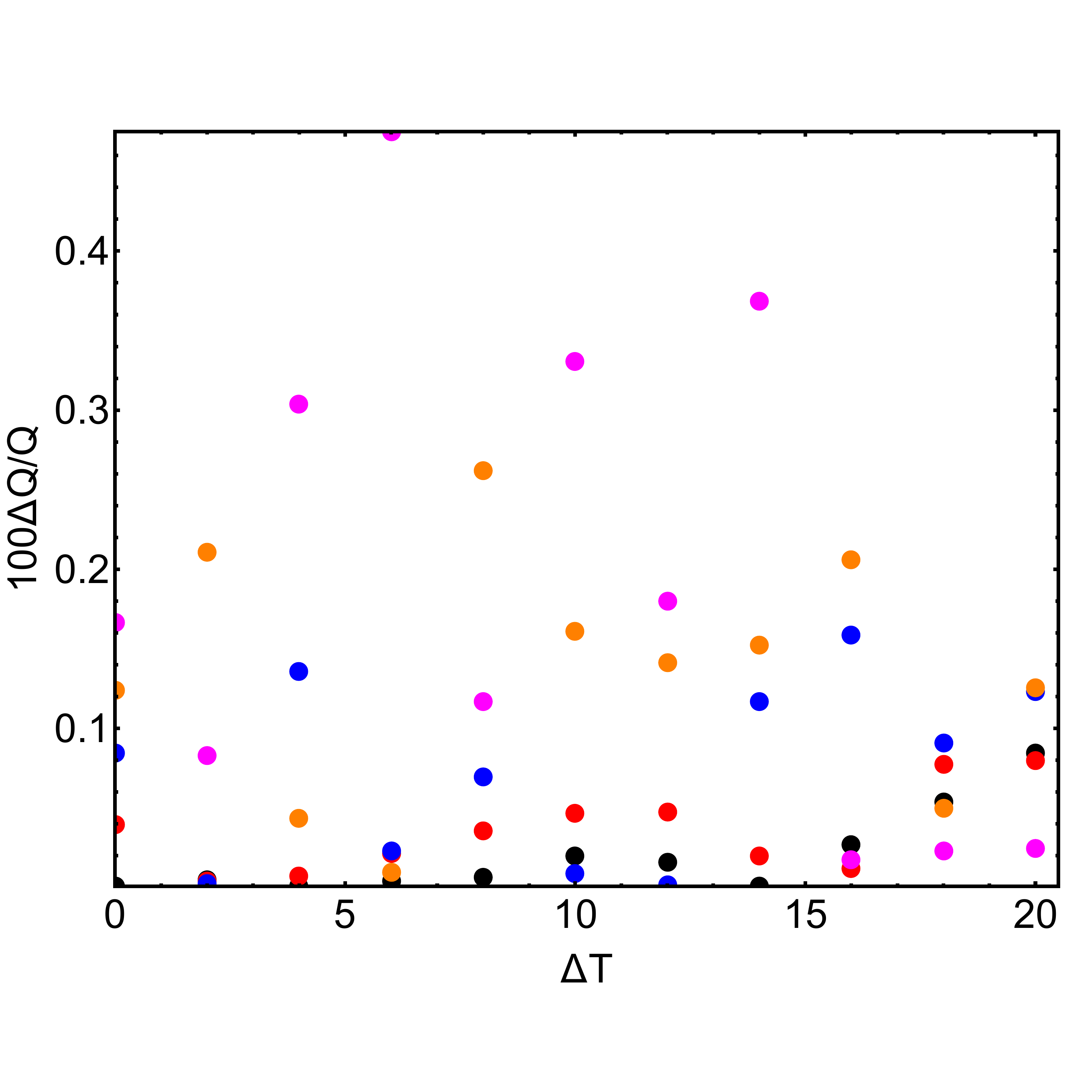}   
\vskip -0.5cm
\caption{\small\it\raggedright Left: Extrapolation of the virial pressure, $Q_v=\lim_{N\rightarrow\infty}Q_{v,N}$ and the pressure  $Q=\lim_{N\rightarrow\infty}Q_{N}$  as a function of $\Delta T$. Big Black, Red, Blue, Orange  and Magenta points are for $Q_v$ and $\gamma=0$, $0.3$, $1$, $5$ and $10$ respectively. Small Gray, Pink, Cyan, Light Orange and Ligth Magenta points are for $Q$ and $\gamma=0$, $0.3$, $1$, $5$ and $10$ respectively. Right: Relative error $100\vert Q-Q_v\vert/Q$. \label{pree3}}
\end{center}
\end{figure}

We see in figure \ref{pree3} (left) the extrapolated $Q$ and $Q_v$. At a glance we do not see any systematic deviation. In fact the relative error $100 \vert Q-Q_v\vert/Q$ runs between $0.01\%$ and $0.4\%$ which is very small and it indicates the good quality of the data obtained in the computer simulation.


\begin{thebibliography}{99}
\bibitem{GL} Garrido, P.L. and Lebowitz J.L. ,{\it Diffusion equations from kinetic models with non-conserved momentum}, Nonlinearity {\bf 31} 54 (2018).
\bibitem{Callen} Callen, H.B., Chapter 17 in {\it Thermodynamics}, J. Wiley and Sons (1960).
\bibitem{S} Spohn, H., {\it Large Scale Dynamics of Interacting Particle Systems}, Springer (1991).
\bibitem{EGLM} Esposito, R., Garrido, P.L., Lebowitz, J.L. and Marra, R., {\it Diffusive limit for a Boltzmann-like equation with non-conserved momentum}, Nonlinearity in press (2019).
\bibitem{delPozo}   del Pozo, J.J., Garrido, P.L. and Hurtado P.I., {\it Scaling laws and bulk-boundary decoupling in heat flow} , Physical Review E {\bf91}, 032116 (2015); {\it Probing local equilibrium in nonequilibrium fluids}, Physical Review E {\bf 92}, 022117 (2015).
\bibitem{Henderson} Henderson, D., {\it A Simple Equation of State for Hard Discs}, Mol. Phys. {\bf 30}, 971 (1975);  {\it Monte carlo and perturbation theory studies of the equation of state of the two-dimensional Lennard-Jones fluid}, Mol. Phys. {\bf 34}, 301 (1977).
\bibitem{DLS} Derrida, B., Lebowitz, J.L. and Speer, E.R., {\it Large Deviation of the Density Profile in the Steady
State of the Open Symmetric Simple Exclusion Process}, Journal of Statistical Physics {\bf 107}, 599 (2002); Bertini, L., Gabrielli, D. and Lebowitz, J.L., {\it Large Deviations for a Stochastic Model of Heat Flow}, Journal of Statistical Physics {\bf 121},  843 (2005).
\bibitem{Bertini} Bertini, L., de Sole, A., Gabrielli, D., Jona Lasinio, G. and Landim, C., {\it Macroscopic fluctuation theory}, Reviews of Modern Physics, {\bf 87}, 593 (2015).
\end{thebibliography}
\end{document}